\documentclass[sigconf, screen]{acmart}
\AtBeginDocument{%
  }

\copyrightyear{2025}
\acmYear{2025}
\renewcommand\footnotetextcopyrightpermission[1]{}  

\acmConference[MM '25]{Make sure to enter the correct conference title from your rights confirmation email}{October 27--31, 2025}{Dublin, Ireland}

\usepackage{algorithm}
\usepackage{algorithmic}
\usepackage{colortbl}
\usepackage{xcolor}
\definecolor{lightcyan}{rgb}{0.88, 1.0, 1.0}
\definecolor{lightblue}{rgb}{.63,.79,.95}
\begin{document}


\title{Breaking the Pre-Planning Barrier: Adaptive Real-Time Coordination of Heterogeneous UAVs}

\author{Yuhan Hu$^*$}
\affiliation{%
  \institution{Sun Yat-sen University}
  \city{Shenzhen}
  \country{China}
}
\email{huyuhan6666@126.com}

\author{Yirong Sun$^*$}
\affiliation{%
  \institution{Digital Twin Institute, Eastern Institute of Technology}
  \city{Ningbo}
  \country{China}
}
\email{win1282467298@gmail.com}

\author{Yanjun Chen}
\affiliation{%
  \institution{Digital Twin Institute, Eastern Institute of Technology}
  \city{Ningbo}
  \country{China}
}
\email{yan-jun.chen@connect.polyu.hk}

\author{Xinghao Chen}
\affiliation{%
  \institution{Digital Twin Institute, Eastern Institute of Technology}
  \city{Ningbo}
  \country{China}
}
\email{xing-hao.chen@connect.polyu.hk}

\author{Xiaoyu Shen}
\affiliation{%
  \institution{Digital Twin Institute, Eastern Institute of Technology}
  \city{Ningbo}
  \country{China}
}
\email{xyshen@eitech.edu.cn}

\author{Wei Zhang$^\dagger$}
\affiliation{%
  \institution{Digital Twin Institute, Eastern Institute of Technology}
  \city{Ningbo}
  \country{China}
}
\email{zhw@eitech.edu.cn}

\renewcommand{\shortauthors}{Hu, Sun and Zhang et al.}


\begin{abstract}
 Unmanned Aerial Vehicles (UAVs) offer significant potential in dynamic, perception-intensive tasks such as search and rescue and environmental monitoring; however, their effectiveness is severely restricted by conventional pre-planned routing methods, which lack the flexibility to respond in real-time to evolving task demands, unexpected disturbances, and localized view limitations in real-world scenarios. To address this fundamental limitation, we introduce a novel multi-agent reinforcement learning framework named \textbf{H}eterogeneous \textbf{G}raph \textbf{A}ttention \textbf{M}ulti-agent Deep Deterministic Policy Gradient (HGAM), uniquely designed to enable adaptive real-time coordination between mission UAVs (MUAVs) and charging UAVs (CUAVs). HGAM specifically addresses the previously unsolved challenge of enabling precise, decentralized continuous-action coordination solely based on local, heterogeneous graph-based observations. Extensive simulations demonstrate that HGAM substantially surpasses existing methods, achieving, for example, a 30\% improvement in data collection coverage and a 20\% increase in charging efficiency, providing crucial insights and foundations for the future deployment of intelligent, flexible UAV networks in complex, dynamic environments.\footnote{Our model and code will be released soon. \\ {* Equal contribution. † Corresponding author: \texttt{zhw@eitech.edu.cn}} }
\end{abstract}




\maketitle

\section{Introduction}
\label{sec:Introduction}
\begin{figure}[t]
    \centering
    \includegraphics[width=1.05\linewidth]{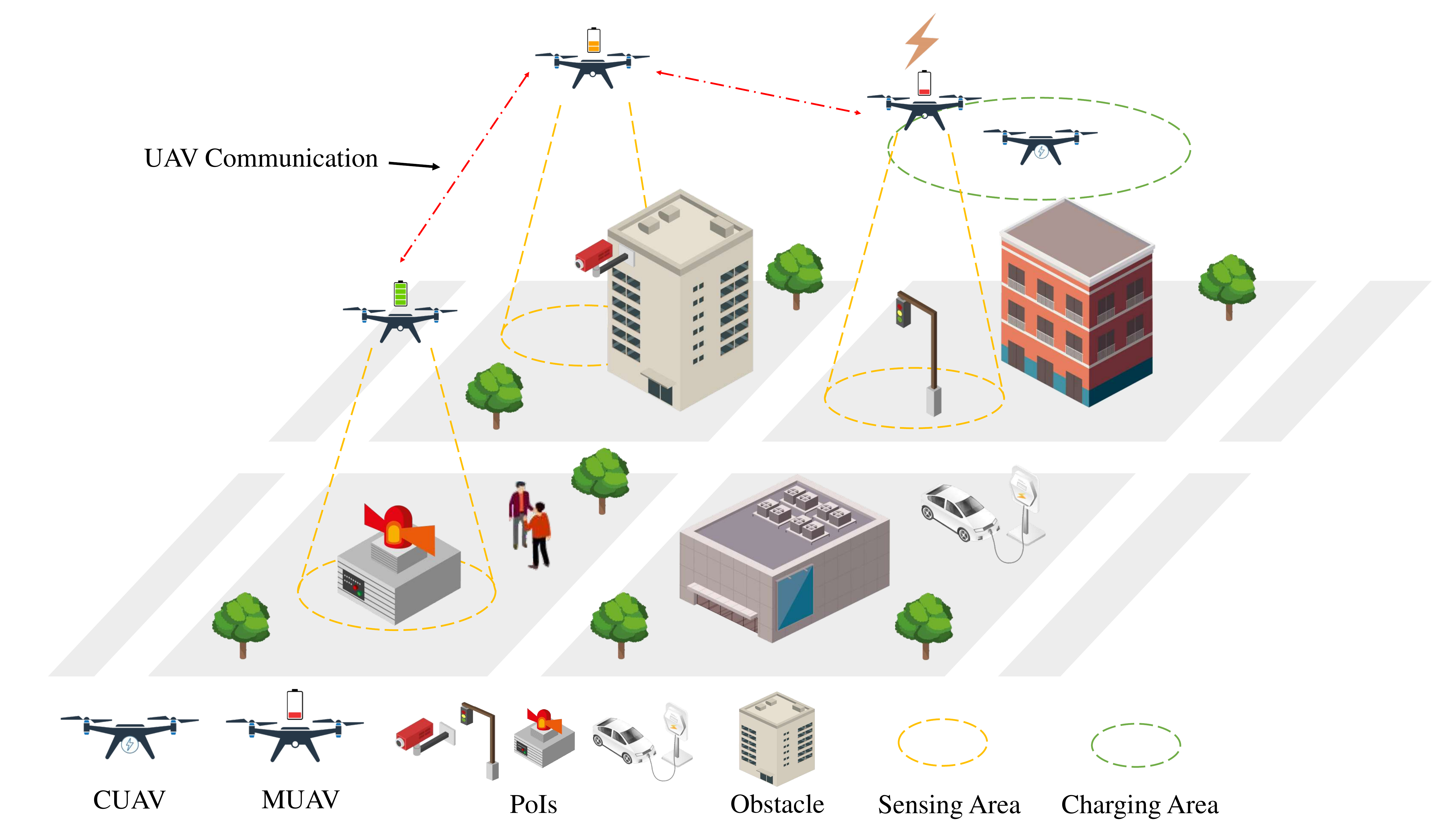}
    \caption{Illustration of adaptive real-time coordination between three MUAVs and a CUAV in a dynamic urban environment. MUAVs autonomously sense and collect data from PoIs, depicted within their sensing range (yellow dashed circles), while CUAV proactively delivers wireless charging to MUAVs in need, indicated by the charging range (green dashed circles). UAV communication (red dashed lines) enables decentralized coordination and obstacle avoidance under limited local observations.}
    \label{fig:enter-label}
\end{figure}
Unmanned Aerial Vehicles (UAVs) have emerged as indispensable tools for executing complex, perception-intensive tasks in dynamic and uncertain environments, including search and rescue, environmental monitoring, and mobile crowd sensing (MCS). Effective deployment of UAVs in such scenarios critically relies on their capability to rapidly adapt trajectories, efficiently avoid obstacles, and continuously sense and collect data from numerous, dynamically evolving points of interest (PoIs). Nevertheless, UAV missions continue to face a fundamental operational bottleneck due to inherent battery limitations, which severely constrain their mission duration and robustness, particularly under urgent or prolonged operational demands.

To alleviate battery constraints, initial efforts have primarily adopted fixed-ground charging stations~\cite{mou2020uav,liu2020energy}, compelling UAVs to periodically interrupt missions and undertake energy-intensive detours for recharging, thereby drastically impairing overall efficiency. To mitigate such inefficiencies, subsequent research explored mobile ground vehicles as dynamic charging platforms~\cite{liu2023dynamic}. More recently, aerial wireless charging approaches, known as "aerial refueling"~\cite{zhu2022aerial,xu2022optimization}, emerged to further minimize mission interruptions by enabling charging UAVs (CUAVs) to recharge mission UAVs (MUAVs) mid-flight. Despite incremental improvements, all these solutions remain fundamentally constrained by their reliance on static, pre-defined routing plans or centralized schedules. Such approaches demand substantial pre-mission planning effort, rendering them inherently inflexible and incapable of responding effectively to unexpected events, evolving task demands, or sudden environmental changes common in realistic deployment scenarios.

To fully realize the potential of UAVs in complex, unpredictable environments, it is therefore essential to transcend traditional static route-planning methods and shift towards truly dynamic, real-time adaptive multi-UAV coordination. Such dynamic coordination, however, poses several intrinsic yet unresolved challenges: (\textbf{i}) UAV trajectories must adapt autonomously and continuously in real-time without pre-planned routes, effectively coping with unexpected environmental changes and unforeseen mission events; (\textbf{ii}) UAV decision-making must rely strictly on decentralized local observations, reflecting realistic operational constraints; (\textbf{iii}) UAV control must operate in continuous action spaces, accurately capturing real-world flight dynamics rather than simplified discrete movements that fail to represent actual UAV maneuverability and precision.

To explicitly address these fundamental challenges, we propose a novel multi-agent deep reinforcement learning framework termed \textbf{H}eterogeneous \textbf{G}raph \textbf{A}ttention \textbf{M}ulti-agent Deep Deterministic Policy Gradient (HGAM). HGAM distinctly integrates heterogeneous graph attention networks (GATs) within a continuous-action actor-critic reinforcement learning architecture to simultaneously and adaptively coordinate MUAVs and CUAVs in real-time, eliminating dependency on predefined trajectories. Specifically, HGAM overcomes challenge (\textbf{i}) by enabling UAVs to continuously adjust flight paths in real-time through dynamic local decision-making; addresses challenge (\textbf{ii}) through an innovative heterogeneous GAT mechanism, which precisely aggregates diverse and locally observed inter-agent information for fully decentralized coordination; and meets challenge (\textbf{iii}) by adopting continuous-action spaces that authentically reflect UAV's maneuverability, enhancing flight precision and operational flexibility. Moreover, advanced training methodologies, further ensure robust and efficient policy learning, enhancing the algorithm's deployability and effectiveness under realistic constraints of partial observability and mission uncertainty.

Extensive simulations validate HGAM's superior adaptive collaboration capability among heterogeneous UAV agents. Results indicate substantial performance improvements over existing methods in critical metrics such as data collection efficiency, geographical fairness, and proactive energy replenishment. Notably, our method consistently maintains mission continuity, promptly reacts to unexpected environmental dynamics, and effectively coordinates MUAVs and CUAVs without predefined routes or obtaining global information. Consequently, this study represents a substantial methodological advancement towards intelligent, autonomous, and practically scalable UAV deployments, significantly expanding the scope and reliability of UAV operations in complex, dynamically evolving environments.    
\section{Related Work}
\label{sec:Related_work}
Energy efficiency and coordination have been critical research challenges for UAV deployment, particularly due to their inherent battery limitations and dynamic operational environments. Existing studies have primarily evolved from fixed-base charging strategies to mobile and aerial recharging solutions, increasingly incorporating sophisticated reinforcement learning and graph-based techniques to enhance flexibility and efficiency.

\subsection{Charging Strategies in UAV Missions}
Early approaches primarily addressed UAV energy constraints through fixed-ground charging stations. ~\cite{mou2020uav} introduced an option-based Deep Q-Network, which effectively enabled UAVs to choose optimal times for recharging at predetermined stations. Similarly, ~\cite{liu2019distributed} leveraged the Ape-X actor-critic framework to enhance UAV path planning for efficient data collection and timely charging~\cite{liu2020energy,fan2022deep}. However, these stationary charging methods inherently required UAVs to divert significantly from their mission paths, increasing travel distance and reducing mission effectiveness. 

To mitigate these inefficiencies, recent research transitioned towards mobile recharging platforms, including ground vehicles~\cite{liu2023dynamic，wang2022mobile} and aerial CUAVs~\cite{xu2022optimization,zhu2022aerial}, aiming to minimize mission interruptions by reducing UAV travel distances. Nonetheless, these methods uniformly rely on pre-defined routes or scheduled coordination sequences, which exhibit inherent drawbacks: route pre-planning processes are time-consuming and resource-intensive, and crucially, pre-planned trajectories lack the flexibility necessary to adapt to real-time changes or unforeseen operational challenges, severely restricting their practicality in dynamic environments.

\subsection{Graph Neural Networks in Multi-UAV Coordination}

Beyond energy-focused strategies, enhancing cooperation among multiple UAVs has motivated integrating Graph Neural Networks (GNNs) into UAV coordination tasks~\cite{zhang2024multi, zhou2022multi}. Graph-based methods have emerged as powerful tools to facilitate multi-agent UAV coordination by explicitly modeling inter-agent interactions and environmental complexity. ~\cite{velivckovic2017graph} have further advanced this direction by dynamically weighting neighbor information, thus enabling decentralized information sharing.

Recent integrations of GAT with reinforcement learning, such as~\cite{dai2020graph, ye2022multi} demonstrated promising results. However, these approaches predominantly rely on discrete action spaces, limiting their maneuverability and responsiveness in highly dynamic environments. Furthermore, they typically assume global observation availability—a condition that rarely holds in practical UAV deployments, where each agent can only perceive its immediate surroundings. Consequently, existing graph-based DRL methods face substantial limitations in scenarios requiring real-time adaptation, fine-grained control, and decentralized decision-making under partial observability.

\subsection{Positioning and Innovation of Our Approach}
Despite significant progress in UAV coordination and energy management, existing studies exhibit critical limitations that impede their practical deployment. Firstly, previous charging strategies, particularly, those utilizing mobile charging platforms—heavily rely on pre-defined trajectories and centralized scheduling. Such methods suffer from inherent inflexibility, as pre-planned paths require extensive planning resources and, crucially, lack the responsiveness necessary to handle dynamic changes or unforeseen events in real-time missions. Secondly, recent graph-based reinforcement learning approaches typically operate with discrete action spaces, restricting the UAVs' maneuverability and fine-grained control. Moreover, these methods usually assume global state observations, an assumption rarely realistic in actual operational scenarios where UAVs inherently have limited, localized perception capabilities.

To address these substantial shortcomings, we propose the \\
\textbf{H}eterogeneous \textbf{G}raph \textbf{A}ttention \textbf{M}ulti-agent Deep Deterministic Policy Gradient. HGAM uniquely embeds heterogeneous graph attention networks within an actor-critic reinforcement learning architecture, explicitly designed to overcome previous methodological constraints. Unlike existing solutions, HGAM requires neither global observation nor pre-defined routes. Instead, it leverages local-field heterogeneous graphs and continuous action spaces, enabling UAVs to dynamically and adaptively coordinate in real-time. Specifically, our heterogeneous GAT mechanism allows UAVs—both MUAVs and CUAVs—to accurately interpret local interaction dynamics, making fully decentralized, fine-grained continuous action decisions to rapidly respond to changing environments and unforeseen operational challenges. To the best of our knowledge, HGAM represents the first method explicitly enabling simultaneous, fully adaptive coordination among heterogeneous UAV teams under continuous action spaces and realistic partial observability conditions, significantly advancing the state-of-the-art beyond previous studiess~\cite{xu2022optimization,zhu2022aerial,dou2024scheduling,chen2022scalable,zhang2022cooperative}.

\section{Problem Formulation}
\label{Problem_formulation}

This section introduces the multi-UAV environment and core notations, defines performance metrics for both MUAVs and CUAVs, and formally states the joint optimization problem under partial observability.

\subsection{System Model}
\label{sec:system_model}

Consider a three-dimensional workspace containing stationary obstacles 
$\mathcal{B}\triangleq\{1,2,\dots,B\}$ and a set of PoIs 
$\mathcal{P}\triangleq\{1,2,\dots,P\}$ randomly distributed across the area. 
We deploy two classes of UAVs: MUAVs $\mathcal{M}\triangleq\{1,2,\dots,M\}$ 
for data collection, and CUAVs $\mathcal{C}\triangleq\{1,2,\dots,C\}$ 
for in-flight recharging of MUAVs. Collectively, all UAVs are represented by 
$\mathcal{U}\triangleq\{1,2,\dots,U\}$, where $U=M+C$.

\begin{figure}[t]
    \centering
    \includegraphics[width=0.7\linewidth]{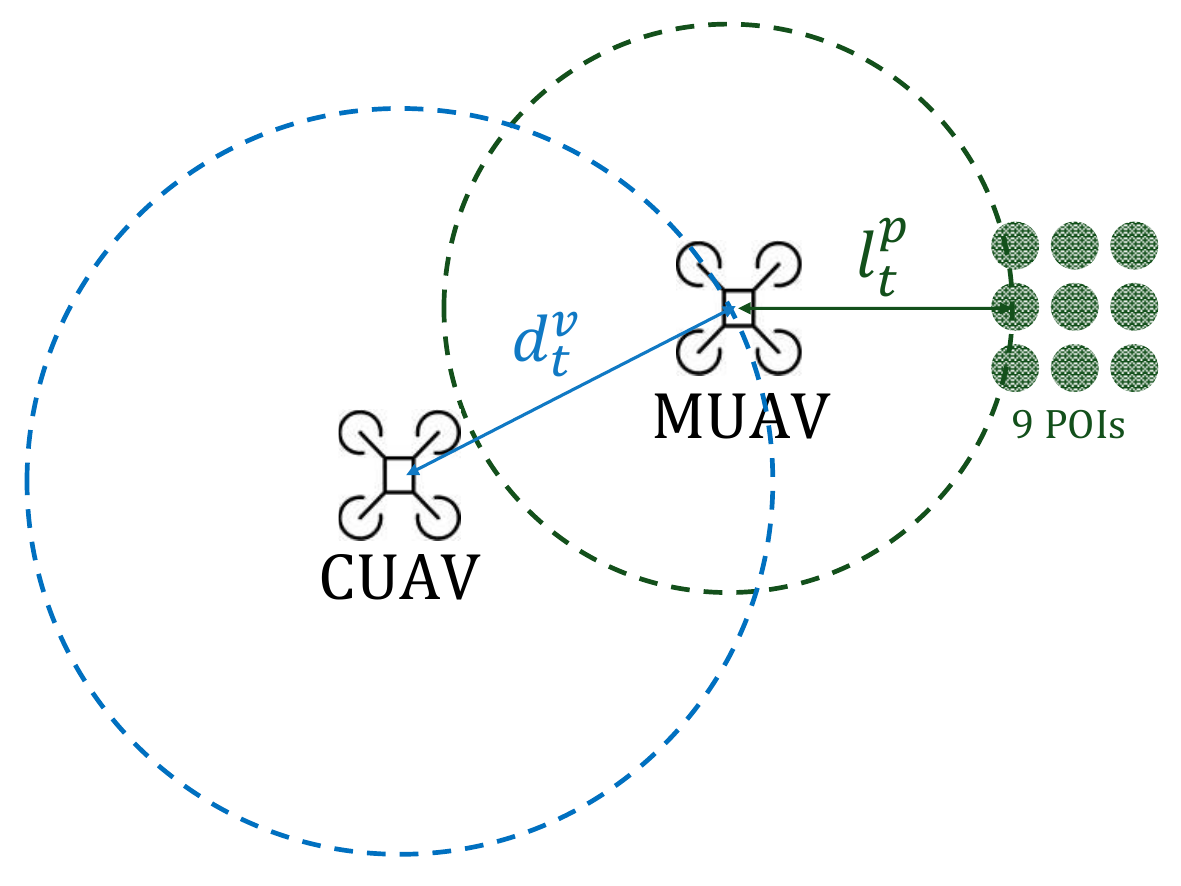}
    \caption{Illustration of MUAV sensing and CUAV charging ranges($d_t^v$ and $l_t^p$), highlighting collaborative interactions with PoIs.}
    \label{fig:uav_area}
\end{figure}

Each MUAV has a sensing range to collect data from nearby PoIs, while each CUAV has a charging radius for wireless energy transfer to MUAVs. To prevent mutual collisions, UAVs operate at different horizontal altitudes, although they may still collide with obstacles or enclosure walls at the same altitude. A global communication link covering the entire workspace allows continuous information exchange among all UAVs.

At the start of each episode, a MUAV $m$ holds a maximum battery level $Er_0^m$, which alone is insufficient for completing the entire mission. The energy consumption at each timestep is modeled as $Ed_t^m \;=\; \beta\,c_t^m \;+\; \kappa\,l_t^m$, where $c_t^m$ is the volume of data collected, $l_t^m$ is the distance traveled, and $\beta,\kappa$ are energy conversion coefficients. Each CUAV provides a constant energy amount $e_0$ per timestep when charging an MUAV. However, only one MUAV can be charged at a time, and if the MUAV’s battery is already full, additional charging is wasted. When multiple MUAVs lie within the CUAV’s charging radius, the CUAV prioritizes the closest MUAV. For main symbol summary used in the system model refer to Appendix~\ref{appendix:System_model}.
 
\subsection{Evaluation Metrics}
\label{sec:evaluation}

We design separate metrics for MUAVs and CUAVs to reflect their respective objectives. MUAVs aim to collect data efficiently and fairly, while CUAVs strive to maintain power support and avoid MUAV depletion.

\paragraph{Data Collection Ratio.}
Let $m_0^p$ be the initial data volume at PoI $p$. Define $D(\pi)$ as the total data volume collected by all MUAVs up to episode $T$. We measure the ratio of collected data to total data:
\begin{equation}
C_T(\pi) \;=\; \frac{D(\pi)}{\sum_{p=1}^P m_0^p}.
\end{equation}

\paragraph{Geographical Fairness.}
To ensure uniform coverage among PoIs, we adopt Jain’s fairness index~\cite{jain1984quantitative}. For PoI $p$, let $\frac{m_T^p}{m_0^p}$ be the fraction of data remaining at $p$. Then
\begin{equation}
\omega_T(\pi)\;=\;\frac{\bigl(\sum_{p=1}^P\,\frac{m_T^p}{m_0^p}\bigr)^{2}}{P\,\sum_{p=1}^P \bigl(\frac{m_T^p}{m_0^p}\bigr)^{2}}.
\end{equation}
Higher $\omega_T(\pi)$ indicates more evenly distributed collection across all PoIs.

\paragraph{Energy Usage Efficiency.}
For each MUAV $m$, let $Ed_T^m$ be total energy consumed, $Er_0^m$ the initial energy, and $Ec_T^m$ the accumulated recharged energy. The overall efficiency is
\begin{equation}
\upsilon_T(\pi)\;=\;\frac{1}{M}\sum_{m=1}^M\frac{Ed_T^m}{\,Er_0^m + Ec_T^m\,}.
\end{equation}

\paragraph{Charging Efficiency.}
In each episode of length $T$, let $T_c$ be the number of timesteps in which CUAV$_c$ is actively charging. We define
\begin{equation}
D_T(\pi)\;=\;\frac{1}{C}\;\sum_{c=1}^{C}\frac{T_c}{T}.
\end{equation}
This indicates the fraction of time that CUAVs collectively spend on effective charging.

\paragraph{Charging Fairness.}
Similarly using Jain’s fairness index, define $E_{\text{max}}$ as the maximum rechargeable energy per MUAV, and $\frac{Ec_T^m}{\,E_{\text{max}}\,}$ the fraction of recharge received by MUAV $m$. We compute
\begin{equation}
F_T(\pi)\;=\;\frac{\bigl(\sum_{m=1}^M\frac{Ec_T^m}{\,E_{\text{max}}\,}\bigr)^{2}}
{M\,\sum_{m=1}^M\bigl(\frac{Ec_T^m}{\,E_{\text{max}}\,}\bigr)^{2}}.
\end{equation}
A higher $F_T(\pi)$ implies a more equitable energy provision among MUAVs.

\subsection{Problem Definition}
\label{sec:problem_definition}

\subsubsection{Objectives and Constraints}
The MUAVs aim to maximize $C_T(\pi)\cdot \omega_T(\pi)$, balancing overall data collection and geographical fairness, while CUAVs seek to maximize $D_T(\pi)\cdot F_T(\pi)$, ensuring efficient and equitable recharging. Formally, the joint objective is:
\begin{equation}
\pi^*\;=\;\arg\max_{\pi}\;\Bigl(\,C_T(\pi)\cdot\omega_T(\pi),\;D_T(\pi)\cdot F_T(\pi)\Bigr),
\label{eq:obj}
\end{equation}
subject to collision avoidance and MUAV energy constraints, i.e. $\forall m\in \mathcal{M},\,Ed_T^m < Er_0^m + Ec_T^m$. 
The episode terminates when a collision occurs or an MUAV’s battery depletes.

\subsubsection{State, Action, and Observation Spaces}
\paragraph{State Space.}
We represent the environment state by the 2D positions of all obstacles, PoIs, UAVs, and relevant energy or data parameters. Each MUAV $m$ tracks $(Er_t^m,Ec_t^m,Ed_t^m)$, while each CUAV $c$ maintains recharging states of MUAVs. We define the system state $s_t\in S$ as a collection of positions, energy levels, and remaining data volumes.

\paragraph{Action Space.}
Each UAV $u$ controls a 2D angular velocity $a_t^u = (x_t^u,y_t^u)\in [-1,1]^2$, normalized so that every UAV moves by the same distance per timestep. MUAVs use these actions to navigate toward PoIs, whereas CUAVs move to charge MUAVs in need.

\paragraph{Observation Space.}
Due to partial observability, each UAV only observes local information within its sensing range (MUAV) or charging radius (CUAV), plus any communicated messages. Specifically:
\begin{itemize}
\item MUAV $m$ observes $o_t^m=\{\mathbf{l}_t,\;\mathbf{b}_t^u,\;\mathbf{p}_t^m,\;v_t^u,\;g_t^u,\;t,\;s_t^u,\;n^u\}$, 
      where $\mathbf{l}_t$ is the set of laser beams measuring distances to obstacles, 
      $\mathbf{b}_t^u$ includes the directions/distances of other UAVs, 
      and $\mathbf{p}_t^m$ describes nearby PoIs.
\item CUAV $c$ observes $o_t^c=\{\mathbf{l}_t,\;\mathbf{b}_t^u,\;\mathbf{e}_t^c,\;v_t^u,\;g_t^u,\;t,\;s_t^u,\;n^u\}$,
      where $\mathbf{e}_t^c$ contains the remaining and charged energy states of MUAVs.
\end{itemize}
These observations are then updated via the observation function \(\Omega(o_{t+1} | s_{t+1}, \mathbf{a_t})\), which reflects the probability of receiving certain partial information given the new environment state $s_{t+1}$.

\subsubsection{State Transition and Reward Functions}
\paragraph{State Transition.}
We denote by $T(s_{t+1}\mid s_t,\mathbf{a}_t)$ the probability that the system transitions from $s_t$ to $s_{t+1}$ after all UAVs execute the joint action $\mathbf{a}_t$. If a collision or MUAV battery depletion occurs, the episode terminates immediately.

\paragraph{Reward Functions.}
Since MUAVs focus on maximizing data collection and fairness, while CUAVs emphasize effective and equitable recharging, we design separate reward structures:
\begin{align}
r_t^m &= h_t^m + \iota_t^m - pl_t^m - pb_t^u, \tag{\text{MUAV reward}} \\
r_t^c &= h_t^c + \iota_t^c - pl_t^c - pb_t^u. \tag{\text{CUAV reward}}
\end{align}
Here,
\begin{itemize}
\item $h_t^m = w_c \times c_t^m$ incentivizes MUAV $m$ to gather more data, 
      while $\iota_t^m$ further encourages discovering or approaching new PoIs.
\item $h_t^c = w_e \times f_t$ rewards CUAV $c$ for effective charging, incorporating a fairness factor $f_t$ that considers both overall charging balance and remaining battery balance among MUAVs (detailed definition provided in Appendix~\ref{appendix:Problem Setting}).
\item $pl_t^m$, $pl_t^c$, and $pb_t^u$ are penalty terms for idle rotation without collecting data, ineffective charging, or collisions/laser beam warnings, respectively.
\end{itemize}

For the CUAV, we define an additional penalty $\iota_t^c$ when it neglects low-battery MUAVs or charges MUAVs that are already sufficiently charged, ensuring the CUAV prioritizes truly urgent charging needs. Moreover, a hierarchical penalty scheme $pl_t^c$ imposes heavier fines when a CUAV chooses suboptimal targets or fails to respond to MUAVs nearing depletion (the explicit formulations of $\iota_t^c$ and $pl_t^c$ are detailed in appendix~\ref{appendix:Problem Setting}). 
Such a design encourages strategic coordination among MUAVs and CUAVs to achieve the dual goals in Eq.~\eqref{eq:obj} while avoiding collisions or mission failures. 

Overall, these definitions incorporate the distinct roles and objectives of MUAVs and CUAVs in a unified multi-agent framework, capturing data collection, fairness, energy efficiency, and safe operations in a single integrated problem.
\section{Proposed Solution HGAM}
\label{sec:proposed_solution_hgam}
This section details our HGAM framework, which incorporates GNN and an actor-critic paradigm to coordinate heterogeneous UAVs under partial observability. We first discuss how to represent UAV states using a heterogeneous graph, then explain the graph feature learning pipeline and actor-critic network architecture, and finally describe the overall training and execution flow.

\subsection{State Representation with a Heterogeneous Graph}
\label{sec:graph_representation}

In our scenario, two types of UAVs---MUAVs and CUAVs---exhibit distinct observation models, reward functions, and objectives, making the environment intrinsically heterogeneous. To accommodate this, we model the multi-agent system as a \emph{heterogeneous} graph $G=(V,E)$. Here, $V$ is the set of node agents (MUAVs and CUAVs), and each node $u\in V$ has a feature vector $v_u$ encoding its local observations (e.g., battery status, position, attribute type). An edge $E(u_1,u_2)=1$ indicates that UAVs $u_1$ and $u_2$ are within communication range and can exchange information in real time. Since UAV positions change over time, these connectivity edges dynamically evolve, making a graph-based approach suitable for capturing agent relationships and topological constraints.

\paragraph{Heterogeneity in Node Features.}
Each node’s feature vector $v_u$ also encodes the agent type (MUAV or CUAV) via a type embedding or attribute flag, ensuring that subsequent network layers can distinguish, for instance, a CUAV’s charging role from an MUAV’s data-collection responsibilities.


\subsection{Graph Feature Learning}
\label{sec:graph_feature_learning}

\begin{figure}[t]
    \centering
    \includegraphics[width=0.45\textwidth]{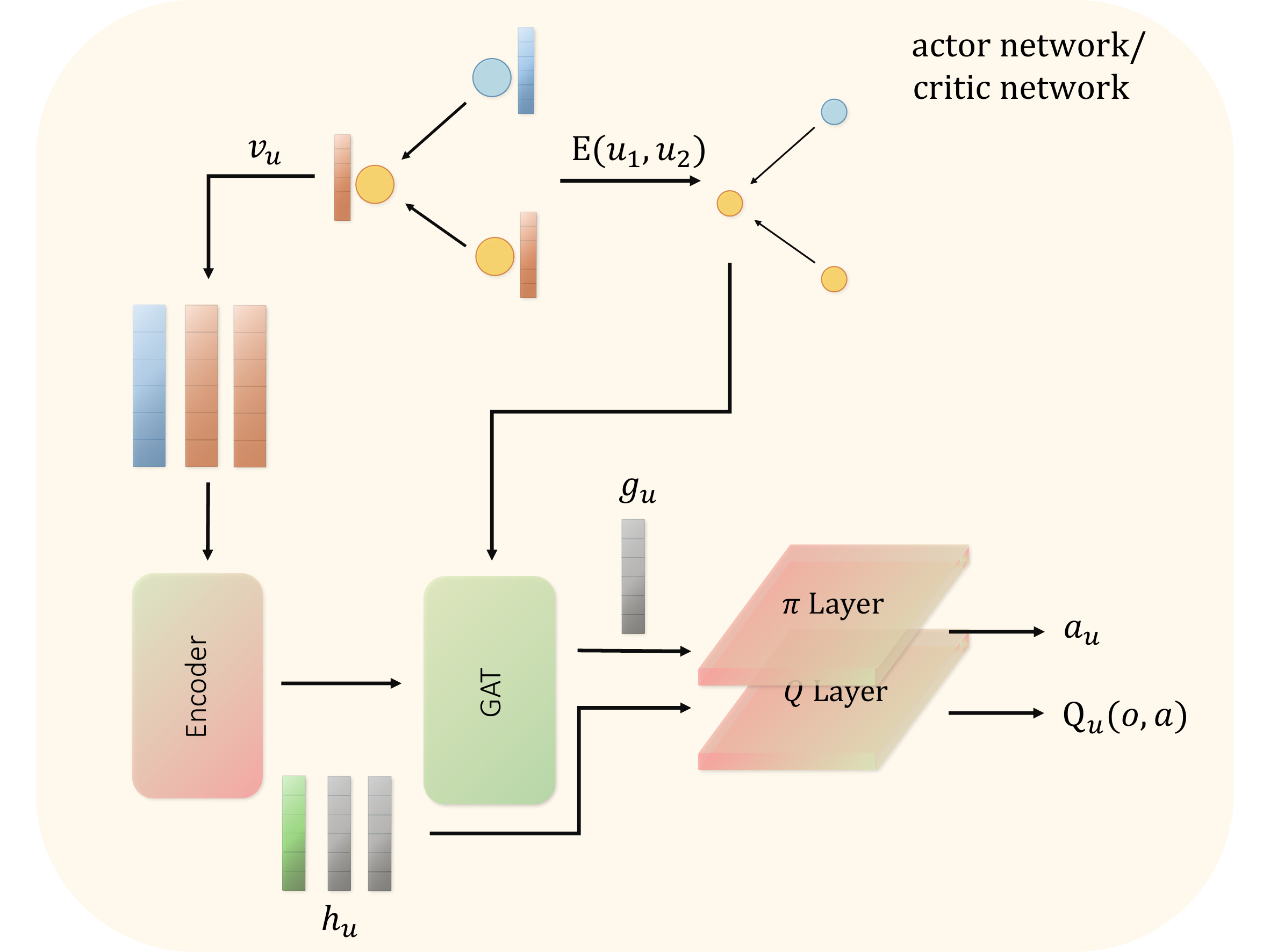}
    \caption{Overview of the actor-critic architecture with heterogeneous GAT. The encoder and GAT module collaboratively generate node embeddings. The actor network ($\pi$ Layer) utilizes local embeddings for decentralized real-time decisions, while the critic network (Q Layer) applies global embeddings for centralized evaluation of joint state-action values, enhancing multi-agent cooperation.}
    \label{fig:ac}
\end{figure}

We design a three-stage pipeline—\emph{encoder}, \emph{GAT layer}, and \emph{execution layer}—to extract informative representations from these heterogeneous graph inputs, as depicted in Figure~\ref{fig:ac}.

\paragraph{Encoder.}
First, each node $u$’s raw feature $v_u$ is processed by an MLP-based encoder $f_u(\cdot)$ to produce an initial embedding $h_u$, i.e.:
\begin{equation}
h_u \;=\; f_u\bigl(v_u\bigr).
\end{equation}
This encoding step unifies variable-dimension observations from MUAVs and CUAVs into a standard embedding dimension, facilitating subsequent attention operations.

\paragraph{Graph Attention Layer.}
Next, each UAV $u$ aggregates information from its neighbors $\mathcal{N}(u)$ via a GAT mechanism~\cite{velivckovic2017graph}. Let $\mathcal{H}(u) = \{h_v\mid v\in\mathcal{N}(u)\}$ be the set of neighbor embeddings. The GAT computes:
\begin{equation}
g_u = t_u\Bigl(h_u,\,\mathcal{H}(u)\Bigr)\;=\;\sum_{v\in\mathcal{N}(u)}\;\alpha_{vu}\,\bigl(W\,h_v\bigr),
\end{equation}
where $W$ is a learnable weight matrix and $\alpha_{vu}$ is an attention coefficient reflecting the relative importance of neighbor $v$ to $u$. Formally,
\begin{equation}
\alpha_{vu} \;=\; \frac{\exp\!\Bigl(\text{LeakyReLU}\bigl(a^\top[\,W\,h_v\,\|\,W\,h_u\,]\bigr)\Bigr)}
{\sum_{k\in\mathcal{N}(u)}\!\exp\!\Bigl(\text{LeakyReLU}\bigl(a^\top[\,W\,h_k\,\|\,W\,h_u\,]\bigr)\Bigr)},
\end{equation}
so that $u$ adaptively focuses on neighbors most relevant for its decision-making. By incorporating agent-type embeddings in $h_u$ and $h_v$, the GAT effectively captures heterogeneous interactions among MUAVs and CUAVs.

\paragraph{Execution Layer.}
Finally, the execution layer combines the node’s own embedding $h_u$ and the GAT output $g_u$ to generate either Q-values (in the critic) or action policies (in the actor). Specifically,
\begin{equation}
Q_u(\mathbf{o},\mathbf{a})\;=\;\psi_u\bigl(h_u^Q,\,g_u^Q\bigr), 
\quad
a_u\;=\;\mu_u\bigl(h_u^\pi,\,g_u^\pi\bigr),
\end{equation}
where $\psi_u(\cdot)$ and $\mu_u(\cdot)$ are MLP heads for critic and actor networks, respectively. Section~\ref{sec:actor_critic} details how these outputs integrate into our Centralized Training and Decentralized Execution(CTDE) framework.

\subsection{Overall Actor-Critic Framework}
\label{sec:actor_critic}
In real-world UAV operations, individual agents operate in a decentralized manner with only local observations, yet effective coordination is essential for mission success. To bridge this gap, we adopt a CTDE strategy. During training, a centralized critic leverages global information to learn a comprehensive Q-function, while each UAV’s actor—operating solely on local data—executes actions in real time, thus aligning with the inherent decentralized nature of UAV deployments.

\paragraph{Local vs.\ Global Graphs.}
During training, the critic constructs a \textit{global} graph, wherein each UAV node $u$ has edges to \emph{all} other nodes, i.e. $N_{\text{global}}(u)=\{v\mid \forall\,v\in V\}$. This holistic view allows the critic to assess the joint state-action value $Q(\mathbf{o},\mathbf{a})$. In contrast, the actor’s \textit{local} graph is restricted to the UAV itself and its closest neighbors of each type, reflecting only partial observations during decentralized execution. Formally, we define
\begin{equation}
\mathcal{N}_{\text{local}}(u)=\{\,v^{(0)},\,v^{(1)}\mid \forall n\in\{0,1\}, d(u,v)<d(u,w)\},
\end{equation}
where $d(u,v)$ denotes the Euclidean distance between UAV $u$ and another UAV $v$. By processing a local subgraph, the actor can operate under real-time constraints without relying on full global state knowledge.

The critic network $\psi_u\bigl(h_u^Q, g_u^Q\bigr)$ evaluates the global Q-value by constructing a global graph that incorporates all UAV observations and actions. Specifically, we define the node feature for UAV $u$ as $v_u = \text{concat}(o_u, a_u)$ to ensure that the critic captures all relevant information from the entire system. This design adheres to the CTDE principle: during training, the critic has access to the full global state, while at execution time, each UAV relies solely on its locally observed data via its actor network. In ideal circumstances, establishing an upper bound on performance, this global view serves as a performance benchmark that decentralized actors can asymptotically approach, even though they operate under more limited, real-time constraints.

\paragraph{Actor Network and Local Graph.}
The actor network $\mu_u(h_u^\pi,g_u^\pi)$ outputs continuous actions $a_u$ based on local embeddings. The node feature of UAV $u$ is $v_u=o_u$, i.e., $u$’s current observation. Together with GAT-aggregated neighbor representations, the actor learns strategies to coordinate with both MUAV and CUAV neighbors, adapting to limited view while collectively maximizing mission objectives.

\paragraph{Critic Network and Global Graph.}
The critic network $\psi_u\bigl(h_u^Q, g_u^Q\bigr)$ evaluates the global Q-value by constructing a global graph that incorporates all UAV observations and actions. Specifically, we define the node feature for UAV $u$ as $v_u = \text{concat}(o_u, a_u)$ to ensure that the critic captures all relevant information from the entire system. This design adheres to the CTDE principle: during training, the critic has access to the full global state—analogous to an offline maximum likelihood estimation~\cite{swamy2025all} that establishes an upper bound on performance—while at execution time, each UAV relies solely on its locally observed data via its actor network. In ideal circumstances, this global view serves as a performance benchmark that decentralized actors can asymptotically approach, even though they operate under more limited, real-time constraints.

\paragraph{Parameter Updates.}
Let $\varphi_u$ and $\theta_u$ denote the parameters of the critic and actor for UAV $u$, respectively. We store agent experiences in a replay buffer $D$, and utilize target networks $\psi_u'$ and $\mu_u'$ for stable updates. The critic’s parameters $\varphi_u$ are updated by minimizing the TD error:
\begin{align}
\label{eq:critic_loss}
\mathcal{L}(\varphi_u)\;=\;\mathbb{E}_{(\mathbf{o},\mathbf{a},\mathbf{r},\mathbf{o'})\sim D}\bigl[\,r_u \;+\;&\gamma\,\psi_u'\bigl(h_u^{Q'},g_u^{Q'};\varphi_u'\bigr)\nonumber\\
&-\;\psi_u\bigl(h_u^Q,g_u^Q;\varphi_u\bigr)\bigr]^{2}
\end{align}
where $h_u^{Q'}$ and $g_u^{Q'}$ are the target embeddings computed from the next-state observations $\mathbf{o'}$ and next actions $\mathbf{a'}$, with $\mathbf{a'}=\mu_u'\bigl(h_u^{\pi'},g_u^{\pi'}\bigr)$.

For the actor, we use a policy gradient that maximizes the critic’s estimated Q-value:
\begin{align}
\label{eq:actor_update}
\nabla_{\theta_u}J(\theta_u)\;=\;\mathbb{E}_{(\mathbf{o},\mathbf{a})\sim D}\Bigl[\nabla_{\theta_u}\,\mu_u\bigl(h_u^\pi,g_u^\pi;\theta_u&\bigr)\,\nabla_{a_u}\,\psi_u\bigl(h_u^Q,g_u^Q\bigr)\nonumber\\
&\Big|_{a_u=\mu_u(h_u^\pi, g_u^\pi  ;\theta_u)}\Bigr]
\end{align}
Here, we backpropagate through the GAT layers and the MLP heads in both actor and critic, ensuring end-to-end learning of graph embeddings tailored to UAV coordination.
\begin{figure}[t]
    \centering
    \includegraphics[width=0.45\textwidth]{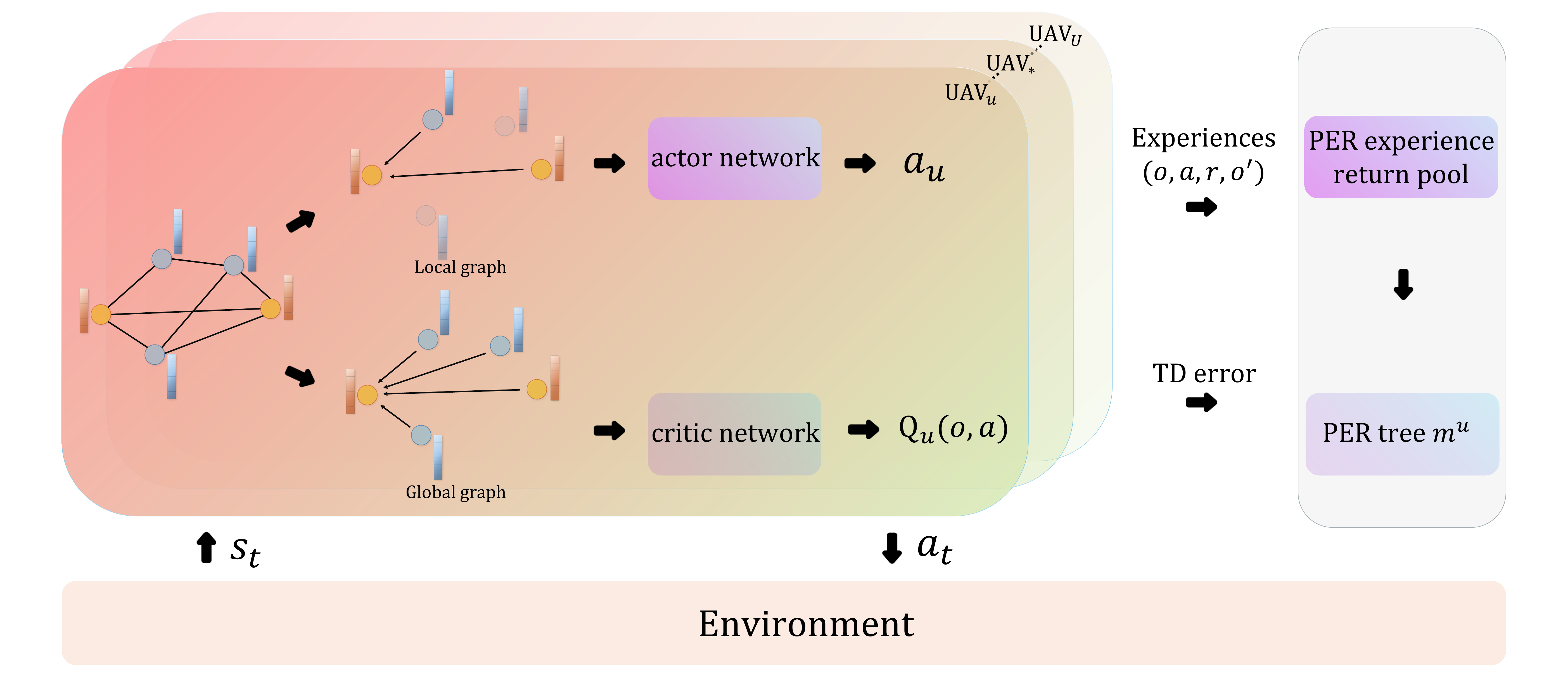} 
    \caption{Overall HGAM pipeline under the CTDE paradigm. Actor networks utilize local graph embeddings for decentralized, real-time decisions, while the critic network employs global graph embeddings for centralized Q-value estimation during training. Experiences collected in the Prioritized Experience Replay (PER) buffer are prioritized based on TD errors, enhancing training stability and performance.}
    \label{fig:Overall_Framework}
\end{figure}

\subsection{Execution and Training Flow}
\label{sec:training_flow}

At runtime (decentralized execution), each UAV only loads its actor network and constructs a local subgraph with neighbors in communication range. The actor computes continuous actions $a_u$ from the local embeddings $h_u^\pi,g_u^\pi$. Periodically, experiences $(\mathbf{o},\mathbf{a},\mathbf{r},\mathbf{o'})$ are stored in the replay buffer. Offline, we conduct centralized training: the critic networks process global observation-action pairs to refine Q-values, and the actor gradients are computed via backpropagation of the TD error. Target networks and soft updates (e.g., $\varphi_u'\leftarrow\tau\varphi_u+(1-\tau)\varphi_u'$) stabilize training.

Overall, HGAM synergizes heterogeneous graph attention with actor-critic to efficiently coordinate MUAVs and CUAVs under partial observability, leveraging local vs.\ global graphs to align with CTDE principles. In the following sections, we demonstrate how this framework improves data collection, charging fairness, and robust multi-UAV coordination.

\section{Training Methodology Design}
\label{sec:Training_methodology_design}

This section details three important strategies we employ to enhance policy convergence and performance under partial observability: (i) a dilemma detection mechanism that prevents MUAVs from falling into local rotation traps, (ii) an N-step return and PER framework to stabilize and accelerate learning, and (iii) an integrated training pipeline under CTDE.

\subsection{Dilemma Detection Mechanism}
\label{subsec:dilemma}

Although MUAVs are designed to navigate toward PoIs for efficient data collection, they can occasionally slip into \emph{local rotation loops}, repeatedly revisiting the same vicinity with limited progress. Inspired by~\cite{wei2022high}, we introduce a detection mechanism to identify and penalize such suboptimal behavior. Specifically, let $o_{t,t+1}$ denote the overlapping area visited by a MUAV between consecutive time steps $t$ and $t+1$. When flying normally, $o_{t,t+1}$ tends to be minimized relative to $o_{t,t'}$ for $t'\neq t+1$, indicating steady movement. However, if there exists a $t'$ such that $o_{t,t'}$\;>\;$o_{t,t+1}$,
the MUAV is likely rotating or circling the same region, signaling a local optimal trap. Once detected, a rotation penalty or modified reward adjustment is applied to discourage such repetitive loops. This ensures MUAVs continually explore or move toward new PoIs rather than wasting time in narrow rotations.

\subsection{N-step Return and Prioritized Experience Replay}
\label{subsec:nstep_per}

Beyond detecting rotation dilemmas, we further boost training efficacy by integrating two well-known reinforcement learning techniques: \emph{N-step returns} and \emph{PER} . Following~\cite{wei2022high}, these improvements address credit assignment challenges and imbalance in experience sampling, especially in multi-agent scenarios.

\paragraph{N-step Return.}
In multi-UAV tasks with delayed rewards (e.g., data collection only becomes meaningful after sufficient travel or charging actions), a longer reward horizon can be crucial. Instead of relying solely on immediate one-step returns, we accumulate rewards over $N$ future steps:
$\lambda_t^u \;=\; r_t^u + \gamma\,r_{t+1}^u + \cdots + \gamma^{N-1}r_{t+N-1}^u,$
where $\gamma\in[0,1)$ is the discount factor. This partial return is then used to compute the target Q-value:
$y_t^u \;=\; \lambda_t^u \;+\;\gamma^N\,\psi_u'\bigl({h_u^{Q'}}_{t}\!,\,{g_u^{Q'}}_{t};\varphi_u'\bigr),$
capturing both short- and mid-term consequences of each agent’s actions.

\paragraph{Prioritized Experience Replay}
Experience replay buffers can become large and diverse. PER~\cite{schaul2015prioritized} ensures that experiences with higher TD errors---indicating more significant learning potential---are sampled more frequently. For each transition $m$, we define its priority based on the TD error 
\(\delta_m^u=y_t^u-\psi_u\bigl(h_u^Q,g_u^Q;\varphi_u\bigr)\). A common weighting scheme is
\begin{equation}
\zeta^u(m)\;=\;\frac{\bigl(\delta_m^u\bigr)^\alpha}{\sum_k\bigl(\delta_k^u\bigr)^\alpha}
\label{eq:3-24-rewrite}
\end{equation}
where $\alpha$ controls how strongly prioritization favors large TD errors. During minibatch sampling, transitions with higher $\zeta^u(m)$ are chosen more often, accelerating the reduction of critical TD errors. Consequently, the critic loss \(\mathcal{L}(\varphi_u)\) is updated as
\begin{align}
\label{eq:3-25-rewrite}
\mathcal{L}(\varphi_u) = \mathbb{E}_{(o,a,r,o')\sim D} \Bigl[\zeta^u(m)\times \Bigl(\lambda_t^u +& \gamma^N\,\psi_u'({h_u^{Q'}}_{t}, {g_u^{Q'}}_{t};\varphi_u') \nonumber \\
& - \psi_u({h_u^Q}_{t}, {g_u^Q}_{t};\varphi_u)\Bigr)^2\Bigr]
\end{align}
Through N-step returns and PER, each UAV’s learning becomes more stable and sample-efficient, key in complex multi-agent environments.

\subsection{Overall Training Process}
\label{subsec:overall_training}

We summarize the integrated training pipeline below. Pseudocode can be found in Appendix~\ref{appendix:train_algorithm}.

\paragraph{Initialization.}
Each UAV $u$ initializes an actor network $\pi_u(o_u;\theta_u)$ and a critic network $Q_u(\mathbf{o},\mathbf{a};\varphi_u)$. MUAVs share a common critic for data collection tasks, whereas CUAVs share another for charging-related objectives. Target networks $\pi_u'$ and $Q_u'$ are cloned from the original networks to stabilize temporal difference learning.

\paragraph{Episode Rollout.}
At the start of each episode, the environment is reset, randomly placing obstacles, PoIs, and UAVs. Each UAV obtains its local observation $o_t^u$. The actor then selects an action $a_t^u\;=\;\pi_u\bigl(o_t^u\bigr)+\mathcal{O}$, where $\mathcal{O}$ denotes Gaussian or Ornstein-Uhlenbeck noise for exploration. UAVs execute actions and receive next observations $o_{t+1}^u$ and rewards $r_t^u$. Each transition $(o_t^u,a_t^u,r_t^u,o_{t+1}^u)$ is stored into a replay buffer $M$, using PER trees $m_u$ to track priorities. If a MUAV enters the local rotation dilemma (Section~\ref{subsec:dilemma}), an additional penalty may be imposed to encourage reorientation.

\paragraph{Batch Sampling and Model Update.}
After accumulating a minimum number of episodes $e_{\min}$, the model begins training while exploration continues:
\begin{enumerate}
\item \textbf{Sample a minibatch} of experiences $H$ from $M$, weighted by PER priorities $\zeta^u(m)$ (Eq.~\ref{eq:3-24-rewrite}).  
\item \textbf{Compute N-step returns:} For each experience in $H$, calculate $\lambda_t^u$ (N-step partial return) and target Q-value $y_t^u$ (Eq.~\ref{eq:3-25-rewrite}).  
\item \textbf{Critic update:} Minimize the TD loss to update $\varphi_u : \varphi_u\;\leftarrow\;\arg\min_{\varphi_u}\;\mathcal{L}(\varphi_u).$
\item \textbf{Actor update:} Maximize the critic-estimated Q-value w.r.t.\ $\theta_u : \theta_u\;\leftarrow\;\theta_u\;+\;\eta\nabla_{\theta_u}\,J(\theta_u)$, where $\nabla_{\theta_u}\,J(\theta_u)$ is computed via Eq.~\ref{eq:actor_update}.
\item \textbf{Target network soft update:} $\varphi_u'\;\leftarrow\;\tau\,\varphi_u+(1-\tau)\,\varphi_u', \theta_u'\;\leftarrow\;\tau\,\theta_u+(1-\tau)\,\theta_u'$.
\item \textbf{Priority update:} Recompute $\delta_m^u$ for each sampled transition and adjust $\zeta^u(m)$ accordingly.
\end{enumerate}
This procedure repeats until collision, battery depletion, or a maximum time horizon is reached, marking the end of an episode. Then a new episode begins.

As training proceeds, MUAVs learn to avoid local rotation dilemmas and effectively collect PoI data, while CUAVs refine their charging policies via N-step returns and prioritized sampling. Empirically, we observe improved stability and faster convergence of the multi-UAV system compared to naive training methods.
\section{Experiment}
\label{sec:experiment}

We evaluate our proposed HGAM approach in a customized multi-UAV environment, comparing it against three baseline methods under both \emph{local view} and \emph{global view} settings. This section details the environment configuration, training hyperparameters, route visualization, and performance results across multiple baselines.

\subsection{Environment Settings}
\label{sec:env_hparam}
All experiments were conducted on an NVIDIA RTX 4090 GPU within a continuous workspace of dimensions $16\times16\times3$ units, representing a realistic operational area rather than a discrete grid. Two MUAVs and one CUAV operate among 100 randomly distributed PoIs with initial data volumes uniformly sampled from [0,1]. Each MUAV has a sensing radius of $1.0$ unit, while the CUAV employs a $1.5$-unit wireless charging radius. UAVs perceive other agents or obstacles within a $4.0$-unit local observation range. Episodes terminate upon collision, battery depletion, or after $700$ timesteps. Detailed hypeparameters, penalty/reward terms, and exact experimental settings are provided in Appendix~\ref{appendix:training_details}.

\subsection{Route Visualization}
\label{sec:route_viz}

\begin{figure}[t]
    \centering
    \includegraphics[width=\linewidth]{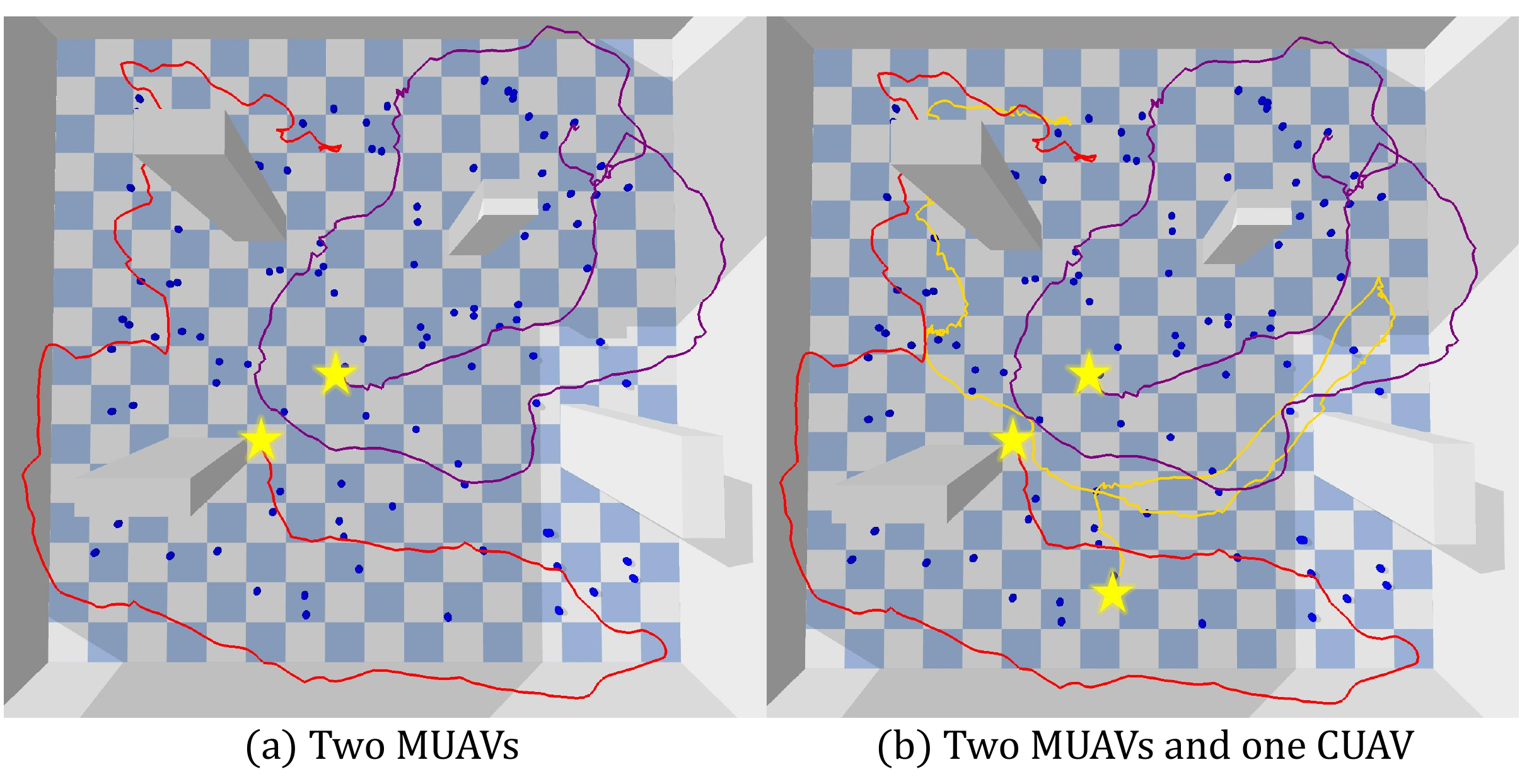}
    \caption{Adaptive UAV trajectories generated by HGAM under local-view constraints. Yellow stars indicate initial positions. (a) MUAV paths (purple and red), demonstrating efficient and complementary coverage. (b) CUAV trajectory (yellow) dynamically supports MUAVs via adaptive charging, while avoiding obstacles.}
    \label{fig:Visualization}
\end{figure}

Before quantitative comparison, we illustrate representative paths taken by two MUAVs and one CUAV. Figure~\ref{fig:Visualization}(a) shows only MUAV trajectories. Despite having only local field-of-view observations, the MUAVs coordinate effectively, covering PoIs in both commonly visited and remote areas, with minimal overlap in their routes. This spatial distribution leads to a high data collection rate.

In Figure~\ref{fig:Visualization}(b), we overlay the CUAV trajectory. Notably, the CUAV (in yellow) initially aligns with MUAV (in purple) and subsequently follows the other MUAV (in red) once it becomes the more urgent charging target. This dynamic following ensures timely wireless recharging for both MUAVs while maintaining collision avoidance with obstacles. Such behavior demonstrates HGAM’s ability to self-organize multi-UAV missions even under partial observability and heterogeneous roles.

\subsection{Baseline Comparison}
\label{sec:comparison}

We benchmark HGAM against three baselines: \emph{Greedy}, \emph{MADDPG}, and \emph{MAAC}~\cite{iqbal2019actor}. 
\begin{itemize}
\item \textbf{Greedy}: A hand-crafted strategy where MUAVs greedily move to the nearest PoIs, and CUAV follows minimal heuristic for charging.
\item \textbf{MADDPG}: A classical multi-agent DDPG framework~\cite{lowe2020multiagentactorcriticmixedcooperativecompetitive} with centralized training, decentralized execution, but lacking explicit graph structures or heterogeneous roles.
\item \textbf{MAAC}: Multi-actor-attention-critic approach, which uses attention in the critic but does not incorporate a heterogeneous GAT-based representation nor distinct local/global graph modeling.
\end{itemize}

We test each approach under two settings:

\paragraph{(1) Local View Training and Evaluation.}
Here, each UAV relies solely on local observations (within its $4.0$-unit communication range) during both training and execution. Table~\ref{tab:local_view_performance} reports MUAV metrics—Data Collection Ratio $(C)$, Geographical Fairness $(\omega)$, Energy Usage Efficiency $(\upsilon)$—and CUAV metrics—Charging Efficiency $(D)$, Charging Fairness $(F)$. HGAM achieves a striking $0.928$ in $C$ (vs.\ $0.630$ for MADDPG and $0.185$ for MAAC) and $0.929$ in $\omega$, showcasing superior coverage and balanced PoI data collection. While $\upsilon$ is slightly lower than MADDPG’s, HGAM still maintains decent energy efficiency $(0.298)$, outstripping Greedy $(0.273)$ and MAAC $(0.042)$. For CUAV-related goals, HGAM obtains $0.613$ in $D$ and $0.969$ in $F$, evidencing equitable and active recharging.

\begin{table}[h!]
\centering
\caption{Comparative evaluation of HGAM against baseline approaches under local-view training and execution conditions, across key performance metrics.}
\label{tab:local_view_performance}
\begin{tabular}{c c c c c}
\toprule
Metric$\uparrow$ & Greedy & MADDPG & MAAC & \textbf{HGAM}\\
\midrule
$C$  & 0.333 & 0.630 & 0.185 & \cellcolor{lightcyan}0.928 \\
$\omega$ & 0.374 & 0.633 & 0.222 & \cellcolor{lightcyan}0.929 \\
$\upsilon$ & 0.273 & 0.333 & 0.042 & 0.298 \\
$D$   & 0.127 & 0.429 & 0.521 & \cellcolor{lightcyan}0.613 \\
$F$   & 0.590 & 0.957 & 0.500 & \cellcolor{lightcyan}0.969 \\
\bottomrule
\end{tabular}
\end{table}

\paragraph{(2) Global View Training and Evaluation.}
To examine robustness, we also train and test the policies with \emph{global} observations, i.e.\ each UAV has full environment visibility. Table~\ref{tab:global_view_performance} shows that HGAM remains superior: $C=0.582$, surpassing MADDPG $(0.492)$ and MAAC $(0.285)$. It likewise leads in $\omega$ $(0.610)$ and $\upsilon$ $(0.422)$. Although the absolute margin is smaller than in the local-view scenario, HGAM retains top-tier performance. For CUAV metrics, $D=0.370$ and $F=0.989$ illustrate HGAM’s robust and consistently high-level charging performance, closely approaching the top-performing baseline. Although slightly behind MAAC in charging fairness $(F=1.000)$, HGAM still demonstrates highly competitive and reliable results.

\begin{table}[h!]
\centering
\caption{Comparative evaluation of HGAM against baseline methods under global-view training and execution scenarios, highlighting the performance across multiple metrics.}
\label{tab:global_view_performance}
\begin{tabular}{c c c c c}
\toprule
Metric$\uparrow$ & Greedy & MADDPG & MAAC & \textbf{HGAM}\\
\midrule
$C$   & 0.333 & 0.492 & 0.285 & \cellcolor{lightblue}0.582 \\
$\omega$ & 0.374 & 0.540 & 0.136 & \cellcolor{lightblue}0.610 \\
$\upsilon$ & 0.273 & 0.305 & 0.023 & \cellcolor{lightblue}0.422 \\
$D$   & 0.127 & 0.403 & 0.000 & 0.370 \\
$F$   & 0.590 & 0.905 & 1.000 & 0.989 \\
\bottomrule
\end{tabular}
\end{table}

\paragraph{Discussion.}
These results indicate that HGAM’s policies are specifically optimized for scenarios involving partially observable environments and decentralized UAV coordination, aligning effectively with realistic and complex operational conditions. The moderate relative performance reduction observed under full observability conditions does not diminish HGAM’s practical value; rather, it underscores the framework’s intentional suitability and robustness for real-world UAV deployments.

\section{Conclusion}
\label{sec:conclusion}
In this paper, we have introduced HGAM, a novel multi-agent deep reinforcement learning framework explicitly developed to address critical limitations of conventional pre-planned routing methods in dynamic, perception-intensive UAV missions. By innovatively embedding heterogeneous graph attention networks within a continuous-action actor-critic architecture, HGAM successfully tackles three fundamental yet previously unresolved challenges: (\textbf{i}) real-time adaptive trajectory adjustments without reliance on predefined routes, (\textbf{ii}) decentralized decision-making under strictly local observations, and (\textbf{iii}) precise maneuverability enabled by continuous action spaces. Extensive simulation results demonstrate that HGAM significantly surpasses existing state-of-the-art approaches in multiple concrete performance metrics, achieving substantially higher data collection efficiency, enhanced geographical fairness, and superior charging coordination effectiveness among MUAVs and CUAVs, even under severe partial observability conditions.

Moreover, HGAM’s capacity to dynamically and autonomously coordinate UAVs positions it as particularly well-suited for realistic and unpredictable operational scenarios, marking a meaningful advancement towards intelligent, flexible, and robust UAV network deployments. Future research will focus on extending the HGAM framework to larger-scale UAV operations, systematically integrating realistic constraints such as sensor noise and communication uncertainties, and rigorously exploring practical deployment challenges through hardware-in-the-loop simulations and real-world experiments, thereby advancing the practical scalability and reliability of autonomous UAV coordination methods.

\newpage


\bibliographystyle{ACM-Reference-Format}
\bibliography{sample-base}
\newpage

\appendix

\section{Main Symbols in System Model}
\label{appendix:System_model}
For clarity and rigorous mathematical treatment of the proposed HGAM framework, we present a systematic summary of the primary symbols and notations used throughout this paper. The symbols defined herein facilitate a consistent and unambiguous description of UAV operations, task objectives, agent states, and multi-agent interactions within our framework. Specifically, these notations include entities representing various environmental components (obstacles, points of interest), UAV categorizations (mission UAVs, charging UAVs), detailed UAV states (energy levels, collected data, movement directions), and core concepts utilized in the reinforcement learning formulation, such as graph representations, policy functions, and priority experience replay indexing. By clearly delineating these elements, Table~\ref{tab:main_symbol} serves as a comprehensive reference, ensuring precise communication of the underlying mathematical structures and enhancing reproducibility and interpretability of our results and analyses.
\begin{table}[htbp]
\scriptsize
\caption{Main Symbol Descriptions}
\label{tab:main_symbol}
\centering
\renewcommand{\arraystretch}{1.0}
\setlength{\tabcolsep}{2.8mm}{
\begin{tabular}{ll}
\toprule[1.2pt]
\textbf{Symbol} & \textbf{Description} \\
\midrule[0.6pt]
$\mathcal{B}, \mathcal{P}$ & Sets of obstacles, PoIs \\
$\mathcal{U}, \mathcal{M}, \mathcal{C}$ & Sets of all UAVs, MUAVs, CUAVs \\
$c_t^m, l_t^m$ & Data collected and distance traveled by MUAV $m$ at time $t$ \\
$m_t^p$ & Remaining data at PoI $p$ \\
$Er_t^m, Ec_t^m, Ed_t^m$ & MUAV energy (remaining, charged, consumed) at time $t$ \\
$d_t^u, l_t^u$ & Direction/distance from UAV $u$ to a target \\
$\mathcal{L}(u)$ & Set of objects in the field of view of UAV $u$ \\
$G=(V,E)$ & Graph representation (nodes, edges) \\
$\mathcal{N}(u)$ & Set of neighbors of $u$ in graph $G$ \\
$n_u$ & PER tree index of UAV $u$ \\
$\pi_u, Q_u$ & Policy and Q-function for UAV $u$ \\
\bottomrule[1.2pt]
\end{tabular}}
\end{table} 

\section{Training Algorithm of Heterogeneous Graph Attention Multi-agent Deep Deterministic Policy Gradient}
\label{appendix:train_algorithm}
For completeness and reproducibility, we provide a detailed step-by-step description of the HGAM training procedure in Algorithm~\ref{alg:training_algorithm}. This algorithm explicitly implements the CTDE paradigm, ensuring effective coordination among heterogeneous UAV agents. Specifically, each UAV possesses an independently operated actor network responsible for real-time, decentralized decision-making based on local observations. Meanwhile, a centrally trained critic network leverages global state-action pairs to accurately estimate the joint value function, guiding individual actors towards cooperative behavior.

The presented algorithm highlights several essential technical components, including PER, which prioritizes sampling experiences with higher significance according to TD error. Such prioritization accelerates the convergence and improves the sample efficiency of the multi-agent reinforcement learning process. Furthermore, the algorithm explicitly integrates a target network soft-update mechanism to enhance training stability, mitigating issues of divergence often encountered in continuous-action reinforcement learning frameworks.

By clearly detailing the initialization, experience collection, prioritized sampling, network updates, and termination criteria, this algorithmic outline provides comprehensive transparency for researchers aiming to implement, validate, or extend the HGAM method in diverse UAV coordination scenarios.

The completed HGAM training algorithm is presented below.
\begin{algorithm}[htbp]
\caption{Training Algorithm of HGAM for $N$ agents}
\label{alg:training_algorithm} 
\begin{algorithmic}[1] 
\STATE Randomly initialize the actor network parameters $\theta_u$ and critic network parameters $\phi_u$ for each UAV, and target actor network parameters $\theta_u'$ and target critic network parameters $\phi_u'$.
\STATE Initialize empty experience replay pool $M$ and PER tree $m_u$ for each UAV.
\FOR{episode $e = 1,2,\ldots,E$}
\STATE Reset the environment, obtain initial $s_0$ and $\textbf{o}_\textbf{0} = (o_0^1, \ldots, o_0^U)$.
\FOR{step $t = 1,2,\ldots,T$}
\STATE Each UAV selects action according to \(a_t^u = \pi_u(o_t^u) = \mu_u(h_u^\pi, g_u^\pi; \theta_u) + \mathcal{O}\)
\STATE Apply all actions $\textbf{a}_\textbf{t} = (a_t^1, \ldots, a_t^U)$ to the environment and obtain the next observation $\textbf{o}_{\textbf{t+1}}$ and reward $\textbf{r}_\textbf{t}$.
\STATE Each UAV updates its own PER tree with fixed upper limit.
\STATE Store experience $(\textbf{o}_\textbf{t}, \textbf{a}_\textbf{t}, \textbf{r}_\textbf{t}, \textbf{o}_{\textbf{t+1}})$ into $M$.
\IF{$e > e_{min}$}
\STATE Sample a batch of experiences $H$ from $M$ using PER tree $m_{t\%U}$ index.
\STATE Calculate $\zeta^u(H)$ with $m_u$ using Eq. 11.
\STATE Update actor network using Eq. 10.
\STATE Update critic network using Eq. 12.
\IF{$e \% f_{soft} == 0$}
\STATE Update target actor and critic networks using \(\varphi_u' = \tau \varphi_u + (1 - \tau) \varphi_u'\) and \(\theta_u' = \tau \theta_u + (1 - \tau) \theta_u'\).
\ENDIF
\STATE Update the priority of experience $H$.
\ENDIF
\IF{UAV collides or MUAV power runs out \textbf{or} reaches maximum time steps}
\STATE Terminate the episode.
\ENDIF
\ENDFOR
\ENDFOR
\end{algorithmic}
\end{algorithm}

\section{Detailed Definitions of Reward and Penalty Functions}
\label{appendix:Problem Setting}
\subsection{Fairness Factor $f_t$}
The fairness factor $f_t$ is computed as a weighted combination of two fairness metrics: the fairness of charged battery levels across MUAVs ($f c_t$) and the fairness of their remaining battery levels ($f r_t$). Specifically, \begin{equation} f_t = w_f \cdot f c_t + (1 - w_f) \cdot f r_t, \end{equation} where $w_f \in [0,1]$ is an adjustable weight parameter that balances these two fairness objectives.

The fairness of the charged battery levels across all MUAVs, $f c_t$, is defined as: \begin{equation} f c_t = \frac{\left(\sum_{i=1}^{n} \min\left(\frac{Ec_t^i}{E_{\text{max}}}, 1\right)\right)^2}{n \cdot \sum_{i=1}^{n} \left[\min\left(\frac{Ec_t^i}{E_{\text{max}}}, 1\right)\right]^2}, \end{equation} where $Ec_t^i$ denotes the charged battery level of MUAV $i$ at time $t$, and $E_{\text{max}}$ represents the maximum battery capacity.

The fairness of the remaining battery levels, $f r_t$, is given by: \begin{equation} f r_t = \frac{\left(\sum_{i=1}^{n} Er_t^i \right)^2}{n \cdot \sum_{i=1}^{n} \left(Er_t^{i}\right)^2}, \end{equation} where $Er_t^i$ represents the remaining battery level of MUAV $i$ at time $t$.

Intuitively, higher values of $f_t$ reflect better fairness in battery levels across all MUAVs, promoting balanced operational longevity and effectiveness.

\subsection{CUAV Penalty for Ineffective Charging $\iota_t^c$}

The penalty term $\iota_t^c$ is designed to penalize CUAV behavior that neglects critical charging opportunities or provides ineffective charging. It is specifically defined as: \begin{equation} \iota_t^c = w_d \cdot l_t^i + w_e \cdot Er_t^i, \end{equation} where:

$i$ represents the MUAV with the lowest remaining battery at timestep $t$;

$l_t^i$ denotes the direct Euclidean distance from the CUAV to MUAV $i$;

$w_d$ and $w_e$ are positive hyperparameters that weight the importance of distance versus battery urgency, respectively.

This penalty structure explicitly encourages CUAV to prioritize moving closer and effectively recharging the MUAV in most urgent need of energy replenishment.

\subsection{Hierarchical Penalty Scheme for CUAV Charging Decisions $pl_t^c$}

To further guide CUAV decisions towards optimal charging strategies, we introduce a hierarchical penalty scheme $pl_t^c$, determined by the relative battery statuses of MUAVs and CUAV actions at each timestep: \begin{equation} 
pl_t^c = 
\begin{cases} 
plow_t^c, & \text{if CUAV is not charging any MUAV,} \\[6pt] \frac{6}{5} \cdot plow_t^c, & \text{if CUAV charges an MUAV already}
\\[4pt] &\quad \text{at maximum battery capacity,} \\[6pt] \frac{plow_t^c}{3}, & \text{if CUAV charges an MUAV whose battery level} \\[4pt] &\quad \text{exceeds the average battery level of all MUAVs,} \\[6pt] \frac{plow_t^c}{4}, & \text{otherwise}. 
\end{cases} 
\end{equation} Here:

$plow_t^c$ represents a baseline penalty applied when the MUAV reaches a critical low battery threshold without receiving timely recharging.

This multi-level penalty scheme incentivizes the CUAV to strategically prioritize urgent charging needs, avoiding ineffective recharging actions that might compromise overall mission objectives and battery fairness among MUAVs.

\section{Training details}
\label{appendix:training_details}
\subsection{Detailed Environment
Settings}
All experiments were conducted on a single NVIDIA RTX 4090 GPU within a simulated continuous workspace of dimensions 16×16×3 units, where the size parameters define the extent of the operational area rather than a discrete grid. In this environment, 100 PoIs are randomly distributed, each initialized with a data volume sampled uniformly from the interval [0,1]. Two MUAVs and one CUAV are deployed to perform data collection and in-flight recharging, respectively.

\subsection{Penalty/Reward Settings}
Each MUAV has a sensing radius of $1.0$ unit for collecting data from nearby PoIs, while the CUAV employs a $1.5$-unit charging radius for wireless energy transfer. Additionally, each UAV can detect other agents or obstacles within a $4.0$-unit field-of-view range. The UAV radius is $0.2$ units, and PoIs have radius $0.1$ units. For each discrete timestep, a UAV travels up to $0.13$ units and can collect up to $0.2$ units of data per hour per PoI. Episodes terminate either upon collision, MUAV battery depletion, or after a maximum of $700$ timesteps.

To encourage proper navigation and charging, we implement various penalty/reward terms. Collisions incur a $100$-point penalty, laser scans below threshold cost $2$ points, and idling MUAVs or those not actively collecting data get penalized. Meanwhile, MUAVs earn a collection reward ($w_c=0.5$) and a small movement reward ($w_l=0.02$), while CUAVs receive $1.6$ points ($w_e$) for effective charging. Additional fairness factors ($w_f=0.5$, etc.) penalize suboptimal or inequitable charging behaviors, ensuring robust multi-agent cooperation. 

\subsection{Hyperparameters}
We employ a Tanh activation in the actor’s final layer (to constrain actions in $[-1,1]$) and LeakyReLU in hidden layers (to preserve negative activation flow). The critic and actor networks have hidden dimensions of $128$ and $64$, respectively, with learning rates $0.001$ (critic) and $0.0001$ (actor). The discount factor $\gamma=0.98$ emphasizes future returns, and the target network soft-update parameter $\tau=0.01$ ensures stable TD learning. An $N$-step return of $3$ is chosen based on preliminary tests for balancing immediate vs.\ delayed rewards. We set the target network update frequency $f_{soft}=50$ and begin training after $e_{min}=50$ episodes. The replay buffer capacity is $100{,}000$ transitions, and the batch size is $128$. In PER, the priority exponent $\alpha=0.6$ controls how strongly TD error affects sampling probabilities.

\subsection{Visualization of model training}
We present the visualizations of data collection rate, episode length, and total reward convergence during model training under both local and global view. For clarity, we focus on the two strongest baselines—MADDPG and MAAC—relative to Greedy. Since MADDPG forms a foundational part of our HGAM model's architecture, comparing their performances in Tables 2 and 3 in the main body of the paper, as well as in Figure \ref{fig:visual_L_merged} and Figure \ref{fig:visual_G_merged}, serves as a partial ablation study.

\paragraph{Visualization of model training - UAVs with local view}

\renewcommand{\thefigure}{6}

\begin{figure*}[t]
    \centering
    \includegraphics[width=0.95\linewidth]{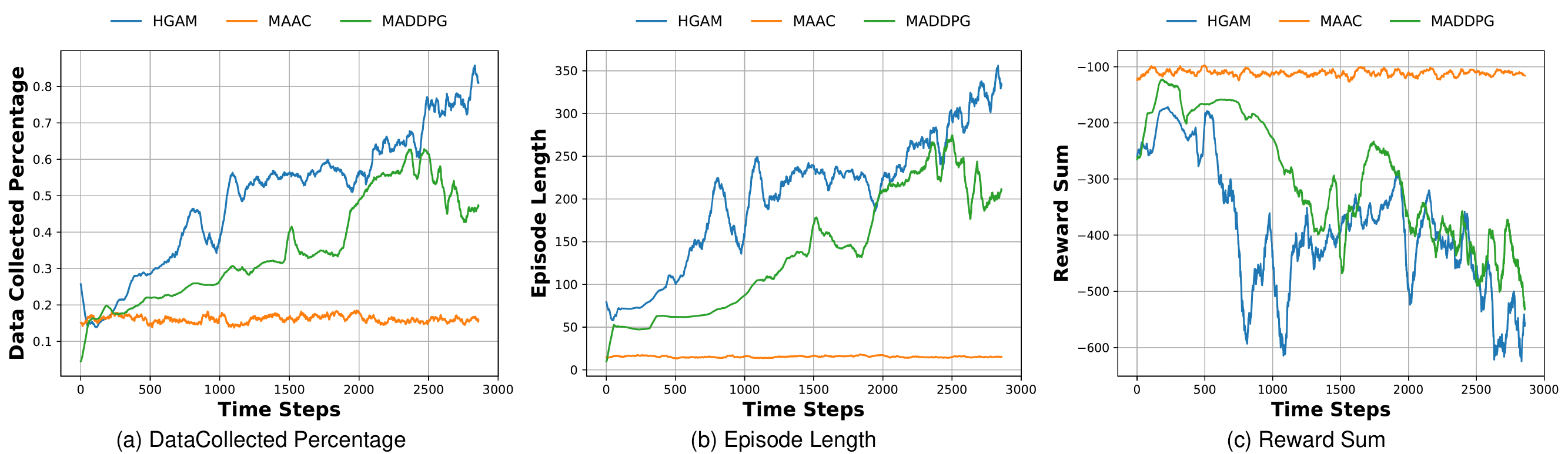}
    \caption{Model training curve visualization comparison: UAVs with local view}
    \label{fig:visual_L_merged}
\end{figure*}

\renewcommand{\thefigure}{\arabic{figure}}

Subplot (a) in Figure \ref{fig:visual_L_merged} shows the percentage of data collected over time for HGAM (blue), MADDPG (green), and MAAC (orange). HGAM outperforms the other models, achieving over 80\% data collection by the end of the training period. This highlights HGAM’s efficiency in data gathering compared to MADDPG, which stabilizes around 60\%, and MAAC, which lags significantly at 20\%. This superior performance suggests that the heterogeneous graphical attention network in HGAM effectively enhances feature extraction from dynamically changing graphs, compensating for information loss due to localized observation.

Subplot (b) in Figure \ref{fig:visual_L_merged} tracks episode length during training. HGAM shows the greatest increase, reaching up to 350 time steps, indicating improved stability and decision-making efficiency. While MADDPG also improves, its episode length increase is less pronounced. MAAC, however, remains nearly constant with short episode lengths, reflecting challenges in sustaining longer episodes, likely due to less effective decision-making.

The reward sum, depicted in Figure \ref{fig:visual_L_merged}(c), illustrates the cumulative rewards over time for the three models. HGAM’s reward sum fluctuates considerably, with an overall downward trend after 2000 time steps, suggesting that while HGAM is actively exploring, it encounters more complex scenarios or suboptimal solutions. MADDPG displays similar fluctuations but to a lesser extent, indicating a more conservative exploration approach. MAAC’s reward sum, in contrast, remains stable and close to zero, indicating minimal learning progress. Despite the similar trend in reward sums between HGAM and MADDPG, HGAM's performance in data collection and geographical fairness in Table 2 exceeds MADDPG by 30\%, showcasing its superior capability.

\paragraph{Visualization of model training - UAVs with global view}
\renewcommand{\thefigure}{7}
\begin{figure*}[t]
    \centering
    \includegraphics[width=0.95\linewidth]{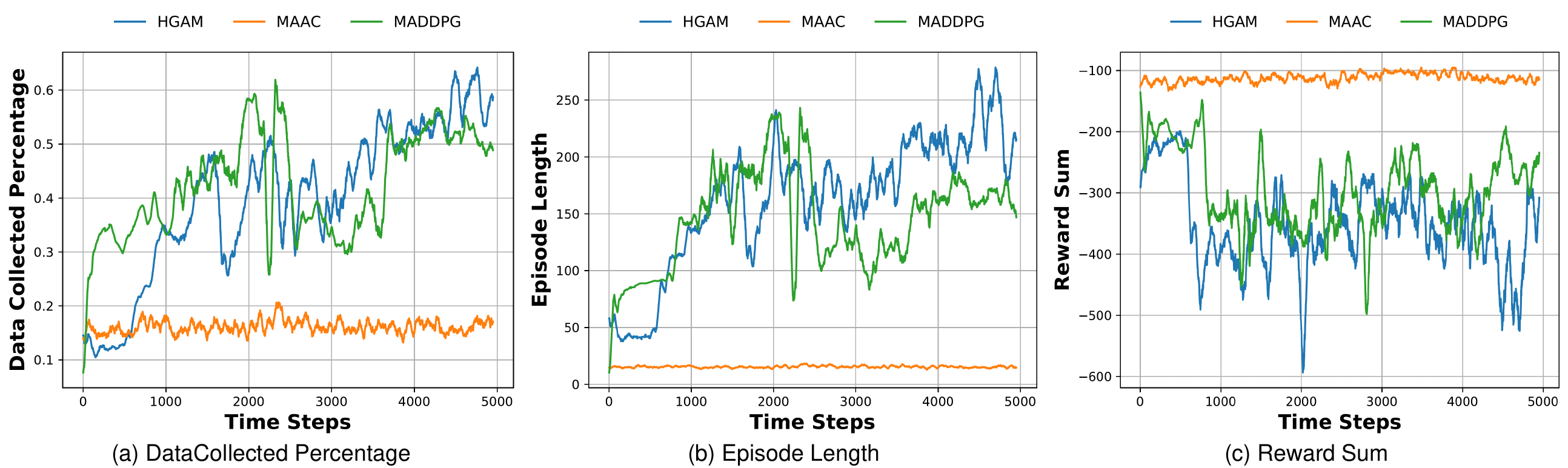}
    \caption{Model training curve visualization comparison:UAVs with global view}
    \label{fig:visual_G_merged}
\end{figure*}
\renewcommand{\thefigure}{\arabic{figure}}

To further demonstrate the robustness of our model, we present visualization figures comparing HGAM with MADDPG and MAAC under a global view.

In Subplot (a) of Figure \ref{fig:visual_G_merged}, we observe the data collection percentage over time for HGAM (blue), MADDPG (green), and MAAC (orange). Initially, MADDPG shows rapid progress, quickly surpassing the other models and reaching approximately 50\% data collection. However, after 2,000 time steps, its performance begins to fluctuate significantly, indicating variability in its effectiveness within the global view. In contrast, HGAM shows a steady and consistent increase, overtaking MADDPG after 3,000 time steps and stabilizing at around 50-55\%. This suggests that HGAM adapts better to the global view, achieving a more reliable and consistent data collection rate. Meanwhile, the MAAC model remains consistently low, around 10-20\%, underscoring its inefficiency in this scenario.

Subplot (b) of Figure \ref{fig:visual_G_merged} illustrates the episode lengths over time. Both HGAM and MADDPG exhibit increasing trends in episode lengths, albeit with significant fluctuations. HGAM shows slightly higher and more consistent episode lengths after 2,000 time steps, stabilizing around 200 time steps towards the end of the training. MADDPG also increases in episode length but with more pronounced fluctuations, suggesting it may be encountering more complex environments or decision-making challenges under the global view. The MAAC model, however, continues to struggle, with episode lengths remaining very short throughout the training, reflecting its poor learning and adaptability.

Finally, Figure \ref{fig:visual_G_merged}(c) displays the reward sum over time. Both HGAM and MADDPG exhibit significant fluctuations in reward sum throughout the training. Although HGAM generally maintains a higher reward sum than MADDPG, both models experience periods of sharp decline, indicating challenges in maintaining consistent performance under the global view. Despite the gap between HGAM and MADDPG narrowing, HGAM still outperforms the other models, as evidenced by the results in Table 3. The reward sum for MAAC remains almost unchanged, with minimal fluctuations and a consistently low reward sum, aligning with its poor performance across the other metrics.

\section{Limitation}
\label{sec:limitation}
Despite the promising results presented in our HGAM framework, several profound limitations and open challenges remain, indicating critical areas for further improvement and exploration.

\subsection{Limitations in Environmental Representation and Perception} Although our approach effectively leverages heterogeneous graph attention and continuous-action multi-agent reinforcement learning to handle partial observability and real-time decision-making, the current method employs a relatively simplified representation of environmental dynamics and uncertainties. Specifically, HGAM utilizes abstracted spatial and positional features to represent the environment, potentially in real-world UAV deployment scenarios where more complex interaction dynamics and various sources of uncertainty may exist. Thus, our method could be enhanced by incorporating more sophisticated models of environmental uncertainty and richer state representations that better reflect realistic operational conditions and multi-agent interactions.

\subsection{Practical Deployment and Scalability Issues} 
Our simulations, while rigorous, remain confined to controlled synthetic environments. Real-world UAV operations inherently involve significantly higher levels of uncertainty, dynamic disturbances, sensor noise, communication latency, and disruptions, none of which are fully captured in synthetic benchmarks. Moreover, traditional evaluation metrics may inadequately capture nuanced aspects of performance in complex, context-sensitive UAV missions. Introducing more advanced and context-aware evaluation methods, potentially involving human-in-the-loop assessment or sophisticated automated benchmarking frameworks, could offer deeper insights into HGAM's true robustness and adaptability. Future work could benefit from extending experiments to more realistic testbeds, including hardware-in-the-loop simulations or actual UAV deployments, to reveal practical constraints and drive further improvements.

\subsection{Intrinsic Algorithmic Robustness and Generalizability} 
The adopted continuous-action reinforcement learning methods (such as MADDPG and its variants) may encounter inherent stability and robustness challenges, particularly in high-dimensional continuous-action spaces. Issues like action distribution distortions, optimization instability, and sample inefficiencies could hinder the method's scalability and generalization to more complex multi-agent tasks. Thus, future research should explore advanced optimization techniques, improved sampling strategies, or corrective mechanisms to further enhance the performance, robustness, and generalizability of HGAM, especially as coordinated UAV missions increase in complexity and scale.
 
\subsection{Challenges in Balancing Local and Global Information}
A significant challenge arises from the trade-off between local decision-making and global mission coordination. Although our CTDE framework successfully leverages local observations for decentralized execution, maintaining coherent global performance becomes increasingly challenging when scaling to larger numbers of agents or broader operational areas. Future research may explore hierarchical or multi-scale reinforcement learning architectures that dynamically balance fine-grained local actions with global strategic oversight, thus ensuring robust collective performance under extreme decentralization and limited communication scenarios.

\end{document}